\documentclass[11pt]{article}

\usepackage[preprint]{acl}

\usepackage{times}
\usepackage{latexsym}

\usepackage[T1]{fontenc}

\usepackage[utf8]{inputenc}

\usepackage{microtype}

\usepackage{inconsolata}

\usepackage{graphicx}
\usepackage{multirow}
\usepackage{colortbl}
\usepackage{tcolorbox}
\usepackage{enumitem} 
\usepackage{algorithm}
\usepackage{algpseudocode}
\usepackage{amsmath}
\usepackage{multirow}
\usepackage[table]{xcolor}
\usepackage{array}
\usepackage{enumitem}
\usepackage{booktabs}
\usepackage{multirow}
\usepackage{amsfonts}
\title{Hard Negative Sampling via Large Language Models for Recommendation}

\author{
\textbf{Chu Zhao}\textsuperscript{1}\quad
\textbf{Enneng Yang}\textsuperscript{1}\quad
\textbf{Yuting Liu}\textsuperscript{1}\quad
\textbf{Jianzhe Zhao}\textsuperscript{1}\quad
\textbf{Guibing Guo}\textsuperscript{1}
\\
\\
\textsuperscript{1}Software College, Northeastern University
\\
{\small\texttt{\{chuzhao, ennengyang, yutingliu\}@stumail.neu.edu.cn}}\\
{\small\texttt{\{zhaojz, guogb\}@swc.neu.edu.cn}}
}

\begin{document}
\maketitle
\begin{abstract}

Hard negative sampling improves recommendation performance by accelerating convergence and sharpening the decision boundary. However, most existing methods rely on heuristic strategies, selecting negatives from a fixed candidate pool. Lacking semantic awareness, these methods often misclassify items that align with users' semantic interests as negatives, resulting in False Hard Negative Samples (FHNS). Such FHNS inject noisy supervision and hinder the model's optimal performance. To address this challenge, we propose \textbf{HNLMRec}, a generative semantic negative sampling framework. Leveraging the semantic reasoning capabilities of Large Language Models (LLMs), HNLMRec directly generates negative samples that are behaviorally distinct yet semantically relevant with respect to user preferences. Furthermore, we integrate collaborative filtering signals into the LLM via supervised fine-tuning, guiding the model to synthesize more reliable and informative hard negatives. Extensive experiments on multiple real-world datasets demonstrate that HNLMRec significantly outperforms traditional methods and LLM-enhanced baselines, while effectively mitigating popularity bias and data sparsity, thereby improving generalization. Implementation code: \url{https://github.com/user683/HNLMRec}


\end{abstract}

\section{Introduction}

Negative sampling \cite{bai2017tapas,ding2020simplify,huang2021mixgcf,zhu2022gain,chen2020efficient,chen2023revisiting} is a key technique in recommender systems. It constructs negative supervision signals to help the model distinguish user-preferred items from non-preferred ones. Recently, hard negative sampling has attracted increasing attention \cite{zhao2023augmented}. It mines more challenging unobserved items and provides stronger gradient signals. This can speed up convergence and sharpen the decision boundary.

\begin{figure}[t] 
    \centering
    \begin{minipage}{0.46\linewidth}
        \centering
        \includegraphics[width=\textwidth]{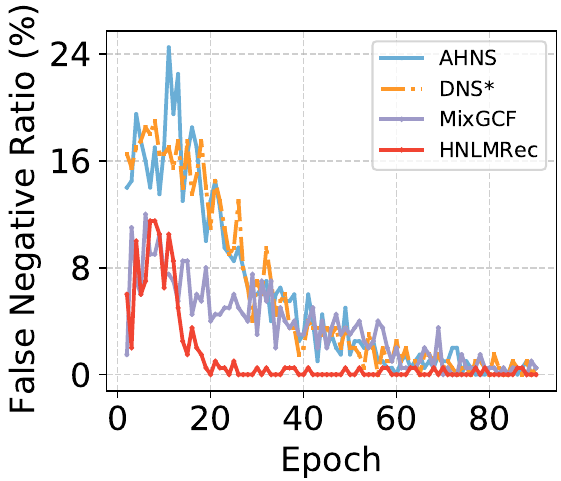}
        \centerline{\small (a)}
    \end{minipage}
    \quad
    \begin{minipage}{0.46\linewidth}
        \centering
        \includegraphics[width=\textwidth]{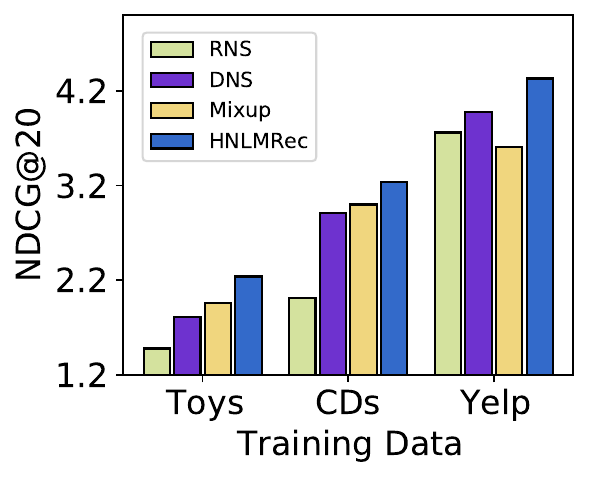}
        \centerline{\small (b)}
    \end{minipage}
    \vspace{-0.4cm}
    \caption{False negative reduction (a) and cross-dataset performance comparison (b).}
    \label{fig: empirical_study}
    \vspace{-0.4cm}
\end{figure}

However, most advanced methods \cite{zhao2023augmented,huang2021mixgcf} rely on ID-based representations and select hard negatives from a fixed candidate pool with heuristic rules. Without semantic awareness, they may mistakenly treat items that are semantically consistent with a user’s interests but structurally distant in the interaction graph as negatives. This produces False Hard Negative Samples (FHNS), injects noisy supervision, and limits the performance upper bound. To study the impact of FHNS, we conduct an empirical analysis of how the FHNS ratio evolves during training and how it correlates with performance. Specifically, we compare the sampled hard negatives with the user's held-out test positives and label a sample as FHNS if it is highly similar to any test positive in embedding space (cosine similarity $\ge \tau$), where $\tau$ is a predefined threshold (set to $0.99$ in our experiments).
As shown in Figure~\ref{fig: empirical_study}(a), all methods exhibit high FHNS ratios early in training, with RNS being the highest. Combined with Figure~\ref{fig: empirical_study}(b), higher FHNS ratios consistently correlate with worse performance across datasets, suggesting that optimizing on FHNS effectively penalizes potential positives, introduces noisy gradients, and degrades both accuracy and generalization.

The emergence of Large Language Models (LLMs) such as GPT have revolutionized multiple domains \cite{wu2023survey,minaee2024large,fei2024multimodal,pang2025language}. With their profound world knowledge and reasoning capabilities, LLMs are theoretically capable of identifying semantically challenging negatives that ID-based methods miss \cite{ren2024representation,wei2024llmrec}.
Nevertheless, simply employing LLMs as "plug-and-play" negative samplers introduces a fundamental dilemma, which we formally define as the "Semantic-Behavioral Gap".
In the semantic manifold of LLMs, items with high linguistic similarity (e.g., \textit{iPhone 13} and \textit{iPhone 14}) are considered "hard" pairs. However, in the collaborative manifold of recommendation, such pairs often exhibit high behavioral correlation (i.e., users upgrading devices), making them positive samples.
Our preliminary analysis reveals that direct LLMs prompting fails to distinguish between \textit{semantic similarity} and \textit{behavioral irrelevance}. Consequently, raw LLM-generated negatives fall into the False Negative trap, introducing severe noise and conflicting gradients that can even untrain the recommender systems. Therefore, how to mitigate the semantic–behavioral gap and directly synthesize hard negatives at the semantic level remains a key challenge.

To bridge this gap, we propose HNLMRec (\textbf{H}ard \textbf{N}egative sampling via \textbf{L}anguage \textbf{M}odel for \textbf{Rec}ommendation), a novel framework for Collaborative-Semantic Alignment in negative sampling. Unlike previous works that treat LLMs merely as black-box generators, HNLMRec theoretically formulates the sampling process as a manifold alignment problem.
Specifically, we move beyond simple prompt engineering and introduce a contrastive supervised fine-tuning mechanism. This mechanism explicitly \textbf{injects} collaborative signals into the LLMs' parameter space. In this way, we constrain the LLM to generate Hard Negative Samples that are \textbf{semantically plausible} (challenging enough to provide informative gradients) but \textbf{behaviorally irrelevant} (distinct enough to be valid negatives).
Furthermore, to tackle the inference latency and text-to-ID mapping issues, HNLMRec synthesizes negatives directly in the embedding space, ensuring seamless integration with existing collaborative filtering  backbones.

Our key contributions are summarized as follows:
\textbf{Methodological Innovation}: We propose HNLMRec, a generative framework that aligns LLMs with collaborative signals. By leveraging contrastive fine-tuning, it effectively filters out false hard negatives while preserving the semantic hardness required for effective training.
\textbf{Theoretical Insight}: We formally identify the \emph{Semantic--Behavioral Gap} in generative negative sampling and explain why naive prompting is sub-optimal due to conflating semantic similarity with behavioral correlation. Moreover, we theoretically show that HNLMRec can retain semantic hardness while suppressing behavioral noise, enabling more stable and less biased optimization.
\textbf{Comprehensive Evaluation}: We conduct extensive experiments on multiple real-world datasets. The results demonstrate that HNLMRec significantly outperforms both state-of-the-art ID-based hard negative samplers and other LLMs-enhanced baselines. 

\section{Preliminary} 
\subsection{Collaborative Filtering}
Given a user set $\mathcal{U} = \{u_i\}_{i=1}^M$ and an item set $\mathcal{V} = \{v_i\}_{i=1}^N$, along with the observed user-item interaction matrix $R \in \mathbb{R}^{M \times N}$, if user $u$ has interacted with item $v$, then $R_{uv} = 1$, otherwise it is $0$. CF methods typically learn an encoder function to map users and items into low-dimensional vector embeddings and predict the user's preference for items by calculating the similarity between these vectors. The Bayesian Personalized Ranking (BPR) \cite{rendle2012bpr} loss function is commonly used to optimize the encoder:
\begin{equation}
    \mathcal{L}_{BPR} = \sum_{(u,v) \in R} -\log(\sigma(s(u,v) - s(u,v^-)))
    \label{eq: BPR_}
\end{equation}
where $s(u,v)$ represents the predicted score of user $u$ for item $v$, $\sigma(\cdot)$ is the sigmoid function and $v^-$ denotes the negatively sampled item. 

\subsection{Negative sampling} The BPR method optimizes the model by contrasting positive and negative samples. Negative samples $v^-$ serve as contrasting signals, allowing the model to discern between items users prefer and those they do not. In-batch random negative sampling is typically used, where items the user has not interacted with are considered negative samples.
High-quality hard negative samples can effectively assist the model in learning better decision boundaries between positive and negative samples. This approach typically involves pre-selecting a candidate set, from which samples are chosen as hard negative samples.

\begin{figure*}[t]
	 \centering
	\begin{minipage}{1\linewidth}
\centerline{\includegraphics[width=\textwidth]{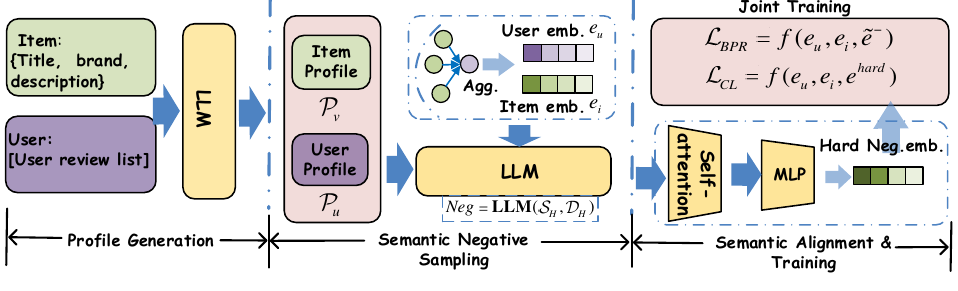}}
	\end{minipage}
    \vspace{-0.4cm}	\caption{The overall pipeline of LLM-driven semantic negative sampling enhancing graph CF consists of three main components: user-item profile generation, semantic negative sampling, and semantic alignment \& training}.
	\label{fig: framwork}
    \vspace{-0.99cm}
\end{figure*}

\section{Methodology}
In Section 3.1 and 3.2, we first introduce the overall HNLMRec framework. In Section 3.3, we provide a theoretical analysis of the bias induced by false hard negatives and explain how our method mitigates it. Finally, in Section 3.4, we discuss the key differences between our approach and self-supervised learning augmentations.
\subsection{The HNLMRec Framework}
\label{sec: methodology}
To extract semantic priors from the LLM, we construct item and user profiles via prompt-based summarization.
For an item $v$ with title $T$, attributes $Atr.$, and reviews $R_v=\{r_1,\ldots,r_n\}$, we form the input prompt $\mathcal{D}_v=\{T,Atr.,R_v\}$ and obtain the item profile by
\begin{equation}
\mathcal{P}_v=\mathbf{LLM}(\mathcal{S}_v,\mathcal{D}_v),
\label{eq: item_profile}
\end{equation}
where $\mathcal{S}_v$ is the item-specific system instruction.
For a user $u$, we build $\mathcal{D}_u=\{R_u,\mathcal{P}_v\}$, where $R_u=\{r_1,\ldots,r_n\}$ are historical reviews and $\mathcal{P}_v$ denotes profiles of interacted items, and derive the user profile by
\begin{equation}
\mathcal{P}_u=\mathbf{LLM}(\mathcal{S}_u,\mathcal{D}_u),
\label{eq: user_profile}
\end{equation}
with $\mathcal{S}_u$ the user-specific system instruction.
Given an interaction pair $(u,v)$, we further construct $\mathcal{D}_H=\{(u,v),\mathcal{P}_u,\mathcal{P}_v\}$ and condition the LLM on $\mathcal{S}_H$ to synthesize semantically plausible yet behaviorally irrelevant candidates.
Detailed prompts for $\mathcal{S}_v$, $\mathcal{S}_u$, and $\mathcal{S}_H$ are provided in the appendix~\ref{prompt}.

\textbf{Generative Hard Negative Sampling}.
Leveraging the constructed profiles, we prompt the LLM to explore the semantic manifold for hard negative candidates.
For a user-item pair $(u,v)$, we design a specific prompt  combined with a hard negative instruction. Crucially, the instruction explicitly specifies the sampling criteria: candidates should be semantically close to the user’s preferences yet behaviorally misaligned with their core interests.
The generation process is formulated as:
\begin{equation}
Neg = \mathbf{LLM}(\mathcal{S}_H, \mathcal{D}_H),
\label{eq: sematic_sampling}
\end{equation}
where $Neg$ denotes the textual representation of the generated candidate. To transform this discrete text into a continuous vector, we employ a pre-trained text embedding model  $\mathcal{T}$ (specifically text-embedding-ada-002 \cite{nussbaum2024nomic}):
\begin{equation}
e^{llm} = \mathcal{T}(Neg).
\label{eq: trans_to_emb}
\end{equation}
We refer to using an LLM to generate a textual negative sample ($Neg$) and then map it to an embedding \emph{without} collaborative-signal SFT as \textit{naive prompting}. The model architecture is illustrated in Figure~\ref{fig: framwork}.

\textbf{Semantic Manifold Alignment and Optimization}.
Since $e^{llm}$ resides in the semantic manifold $\mathcal{M}_s$ (with high dimensionality, e.g., 4096), it cannot be directly interacted with the collaborative manifold $\mathcal{M}_c$. To bridge this gap, we introduce a \emph{Semantic Alignment Transformer} to first contextualize and align semantic representations via self-attention, followed by a learnable Manifold Projector parameterized by a non-linear MLP. Concretely, we apply self-attention to obtain an aligned semantic embedding:
\begin{equation}
\hat{e}^{llm} = \mathrm{Self-Attention}(e^{llm}),
\end{equation}
where $\mathrm{Self-Attention}(\cdot)$ denotes a (lightweight) self-attention module that captures intra-semantic dependencies and produces a task-aware aligned representation. Then, the aligned embedding is projected onto the collaborative latent space:
\begin{equation}
e^{hard} = W_2 \cdot \text{ReLU}(W_1 \cdot \hat{e}^{llm} + b_1) + b_2,
\label{eq: map}
\end{equation}
where $W_1, W_2, b_1, b_2$ are learnable parameters. The resulting $e^{hard}$ represents the final hard negative embedding aligned with the recommendation task.
The primary recommendation objective is optimized via BPR loss:
\begin{equation}
\mathcal{L}_{\mathrm{BPR}}
= - \sum_{(u, v^{+}) \in \mathcal{G}}
\log \sigma\!\left( \mathbf{e}_{u}^{\top}\mathbf{e}_{v^{+}} - \mathbf{e}_{u}^{\top}\mathbf{e}^{\mathrm{hard}} \right).
\label{eq: BPR}
\end{equation}
where $\mathbf{e}_{u}, \mathbf{e}_{v^{+}}$ are embeddings from the backbone CF model (e.g., LightGCN \cite{he2020lightgcn}). $\mathcal{G}$ denotes the user--item bipartite interaction graph constructed from the interaction matrix $\mathbf{R}$.
To explicitly inject collaborative signals into the projection process and suppress False Hard Negatives (FHNS), we propose a \emph{Semantic--Collaborative Alignment} objective.
Specifically, we align the projected hard negative embedding $e^{hard}$ with a \emph{collaborative hard negative label} mined by the backbone CF model, and contrast it against other in-batch items via an InfoNCE loss~\cite{wu2021rethinking}:
\begin{equation}
\label{eq: align}
\mathcal{L}_{\text{align}}
= - \frac{1}{|\mathcal{B}|}
\sum_{(u, v^+, v^-_{cf}) \in \mathcal{B}}
\log
\frac{
\exp\!\left(s_{v^-_{cf}} / \tau_c\right)
}{
\sum_{v \in \mathcal{V}_{\mathcal{B}}}
\exp\!\left(s_{v} / \tau_c\right)
},
\end{equation}
where $\mathcal{B}$ is a mini-batch of triples $(u,v^+,v^-_{cf})$ with user $u$, positive item $v^+$, and CF-mined hard negative item $v^-_{cf}$ (used as the positive target in InfoNCE). We define $s_v := \kappa(e^{\text{hard}}, e_v)$, where $e_v$ is the embedding of item $v$ in the collaborative space, $\kappa(\cdot,\cdot)$ is a similarity function, $\tau_c$ is the contrastive temperature, and $\mathcal{V}_{\mathcal{B}}$ denotes the in-batch item set used
in the denominator as negatives.

This objective encourages the projector to preserve semantic hardness while grounding it in collaborative behaviors, thereby filtering out semantically similar but behaviorally positive FHNS.
The overall objective function is a multi-task learning formulation:
\begin{equation}
\mathcal{L}_{\Theta} = \mathcal{L}_{BPR} + \lambda_1 \cdot \mathcal{L}_{align} + \lambda_2 \cdot | \Theta |^2_F,
\end{equation}
where $\lambda_1$ and $\lambda_2$  are hyperparameters. The architecture of the entire pipeline is illustrated in Figure \ref{fig: framwork}, and the training procedure is detailed in Algorithm \ref{alg: llm-driven-sampling}.
In Section~\ref{sec: theory_gap_and_justification} (Part 2), we provide a theoretical analysis showing \textbf{ why naive prompting strategy produces FHNS and consequently limits model performance}. The results in Table~\ref{tab: empirical-toys-cds} for the \textit{LLMs w/o Alignment} setting further corroborate our theoretical findings.

\subsection{Collaborative-Semantic Alignment via Fine-tuning in HNLMRec}

\textbf{Hybrid Prompt Construction}. To incorporate behavioral constraints into the LLM input, we construct \emph{Hybrid Prompts} that fuse collaborative embeddings with semantic profiles.
We use a pre-trained CF model (e.g., MixGCF~\cite{huang2021mixgcf}) to obtain collaborative embeddings $\{\mathbf{e}_u, \mathbf{e}_v\}$.
Since $\mathbf{e}_u,\mathbf{e}_v$ lie in a low-dimensional collaborative space, we introduce an MLP-based projector $\psi(\cdot)$ to map them into the LLM representation space.
The final query $\mathcal{Q}$ concatenates the task instruction $\mathcal{I}$, projected behavioral signals, and semantic profiles $\{\mathcal{P}_u,\mathcal{P}_v\}$:
\begin{equation}
\mathcal{Q} = \text{Concat}\big(\mathcal{I}, \psi(\mathbf{e}_u), \mathcal{P}_u, \psi(\mathbf{e}_v), \mathcal{P}_v\big).
\label{eq: prompt_const}
\end{equation}
This construction conditions the LLM on \emph{both} semantic context and historical interaction patterns, providing an explicit behavioral grounding beyond pure text prompting.

\textbf{Contrastive Supervised Fine-tuning}. To eliminate semantic shift and hallucinations, we fine-tune the LLM to directly synthesize hard negative representations in the embedding space rather than generating free-form text \cite{wang2023improving}. After SFT, Eq.~\ref{eq: sematic_sampling} and ~\ref{eq: trans_to_emb} are replaced by extracting $e_{\mathrm{llm}}$ from the LLM's last-layer hidden state; the external embedder $T$ is only used in the \emph{naive prompting} baseline.
We treat the hard negatives produced by the pre-trained CF model as "Collaborative Labels" to supervise the LLM's output. By appending an [EOS] token to both the hybrid prompt and the target label, we extract the hidden states of the last layer to obtain the query embedding $\mathbf{Q}_{emb}$ and the label embedding $\mathbf{N}^+_{emb}$. We denote by $s^+$ the alignment score between the query representation and its target collaborative label, and by $s_i^-$ the score between the query and the $i$-th in-batch negative sample:
$
s^+ = \mathbf{Q}_{\mathrm{emb}}^\top \mathbf{N}^{+}_{\mathrm{emb}},
s_i^- = \mathbf{Q}_{\mathrm{emb}}^\top \mathbf{N}^{i}_{\mathrm{neg}} .
$
With these definitions, the InfoNCE-style supervised fine-tuning (SFT) objective is shown as:
\begin{equation}
\label{eq:SFT_refined}
\mathcal{L}_{\mathrm{SFT}}
= -\log
\frac{\exp\!\left(s^+/\tau_s\right)}
{\exp\!\left(s^+/\tau_s\right) + \sum_{i=1}^{N} \exp\!\left(s_i^-/\tau_s\right)}.
\end{equation}
where $\mathbf{N}_{neg}^i$ denotes in-batch negative samples. We employ LoRA \cite{hu2021lora} for parameter-efficient adaptation. Through this process, the LLM learns to distill its world knowledge into collaborative-aware embeddings, producing Hard Negatives that are semantically challenging yet behaviorally distinct. The synthesized embeddings $e^{llm}$ are then fed into the downstream CF training pipeline.

\subsection{Theoretical Analysis and Justification: Bridging the Semantic--Behavior Gap}
\label{sec: theory_gap_and_justification}
Section 3.3 formalizes the negative sampling task and the ideal target distribution $q^\star$. We show that naive LLM prompting without behavioral constraints tends to sample FHNS that overlap with users’ positives, and quantify their mass by the noise rate $\pi(u, v^+)$. We then analyze how FHNS introduce sampling-induced gradient bias via a mixture decomposition, causing direction conflicts with bias that scales with $\pi(u, v^+)$. Finally, we justify HNLMRec from conditional probability shift and mutual-information maximization: injecting collaborative signals yields a behavior-conditioned distribution $q_{\text{aligned}}$ that reduces $\pi_{\text{aligned}}$, while InfoNCE-based alignment maps semantics to collaboration and mitigates hallucination, asymptotically eliminating the bias as $\pi_{\text{aligned}}\to 0$.

\paragraph{1) Task and the ideal target sampling distribution.}
Let $\mathcal{V}$ denote the item universe, and let $\mathcal{I}_u^+ \subset \mathcal{V}$ denote the set of observed positive items (interactions) of user $u$.
Given an observed positive item $v^+ \in \mathcal{I}_u^+$, the goal of negative sampling is to draw \emph{true hard negatives} that are \emph{semantically plausible} yet \emph{behaviorally irrelevant}.
Let $d_S(\cdot,\cdot)$ be a semantic distance and $\mathcal{N}_S(v^+)$ the semantic neighborhood of $v^+$.
The ideal support set is
\begin{equation}
\mathcal{H}_u(v^+) \;=\; \mathcal{N}_S(v^+) \setminus \mathcal{I}_u^+.
\end{equation}
Accordingly, the ideal (behavior-constrained) target distribution is
\begin{equation}
\begin{aligned}
 q^\star(v\mid u,v^+) \;= & \frac{1}{Z^\star(u,v^+)}
\exp\!\big(-d_S(v,v^+)/\tau\big) \\ & \cdot \mathbb{I}\{v\notin \mathcal{I}_u^+\},
\label{eq:q_star}
\end{aligned}
\end{equation}
where $Z^\star(u,v^+)$ is the normalizer. Thus, $q^\star$ concentrates probability mass on semantic neighbors of $v^+$ that are \emph{not} positives for $u$.

\paragraph{2) Why naive prompting yields FHNS.}
A naive LLM prompting strategy is dominated by the pretrained language prior and typically samples from the semantic neighborhood \emph{without} enforcing behavioral constraints:
\begin{equation}
\begin{aligned}
 q_0(v\mid v^+) \; &=  \frac{1}{Z_0(v^+)}
\exp\!\big(-d_S(v,v^+)/\tau\big) \\ &\cdot \mathbb{I}\{v\in \mathcal{N}_S(v^+)\},
\label{eq:q0}
\end{aligned}
\end{equation}
where $Z_0(v^+)$ is the normalizer. Note that $q_0(\cdot\mid v^+)$ does not depend on $u$; however, once we evaluate samples against the user-specific positives $\mathcal{I}_u^+$, $q_0$ induces a user-dependent false/true negative mixture.
Define FHNS as the overlap between semantic neighbors and user positives:
\begin{equation}
\mathcal{V}_{\mathrm{FHN}}(u,v^+) \;=\; \mathcal{N}_S(v^+)\cap \mathcal{I}_u^+.
\end{equation}
The FHN noise rate (probability mass of FHNS under $q_0$) is
\begin{equation}
\begin{aligned}
 \pi(u,v^+) \;=\; & \mathbb{P}_{v\sim q_0(\cdot\mid v^+)}[v\in \mathcal{I}_u^+]
\;= \\ &\; \sum_{v\in \mathcal{V}_{\mathrm{FHN}}(u,v^+)} q_0(v\mid v^+).
\label{eq:fhn_mass}
\end{aligned}
\end{equation}
In recommendation domains, semantic proximity exhibits a strong \emph{enrichment} effect for behavioral relevance, implying that $\pi(u,v^+)$ is substantially larger than the noise level of random sampling (e.g., $\tfrac{|\mathcal{I}_u^+|}{|\mathcal{V}|}$). This effect is often amplified for popular items due to the overlap between LLM pretraining bias and interaction/exposure bias.

\paragraph{3) Gradient bias induced by FHNS.}
Consider a pairwise loss
\begin{equation}
\ell(u,v^+,v^-)\;=\;\phi\!\big(s_{u,v^-}-s_{u,v^+}\big),
\end{equation}
where $s_{u,v}$ is the model score and $\phi(\cdot)$ is a monotonically increasing surrogate.
Define the sampling-induced gradient bias as the gap between expected gradients under $q_0$ and $q^\star$:
\begin{equation}
\label{eq:grad_bias_def}
\begin{aligned}
\mathrm{Bias}(u,v^+)
& = 
\mathbb{E}_{v^-\sim q_0(\cdot\mid v^+)}
\!\left[\nabla_\theta \ell\right]
\\ & -
\mathbb{E}_{v^-\sim q^\star(\cdot\mid u,v^+)}
\!\left[\nabla_\theta \ell\right].
\end{aligned}
\end{equation}
Using the law of total probability over the event $\{v\in\mathcal{I}_u^+\}$, decompose $q_0$ as
\begin{equation}
\begin{aligned}
q_0(\cdot\mid v^+) \; & =  \; \pi(u,v^+)\, q_{\mathrm{FHN}}(\cdot\mid u,v^+) \\ &+ \big(1-\pi(u,v^+)\big)\, q_{\mathrm{TN}}(\cdot\mid u,v^+),
\label{eq:q0_mixture}
\end{aligned}
\end{equation}
where $q_{\mathrm{FHN}}$ and $q_{\mathrm{TN}}$ are $q_0$ conditioned on $v\in\mathcal{I}_u^+$ and $v\notin\mathcal{I}_u^+$, respectively.
Focusing on the dominant FHN-induced distortion yields
\begin{equation}
\begin{aligned}
\mathrm{Bias}(u,v^+)
 & \approx 
\pi(u,v^+)\Big(
\mathbb{E}_{v\sim q_{\mathrm{FHN}}}\!\left[\nabla_\theta \ell\right] \\ &
- 
\mathbb{E}_{v\sim q_{\mathrm{TN}}}\!\left[\nabla_\theta \ell\right]
\Big).
\label{eq:grad_bias_decomp}
\end{aligned}
\end{equation}

\smallskip
\noindent\textbf{Key implications.}
(i) \emph{Directional conflict:} for $v\in\mathcal{V}_{\mathrm{FHN}}(u,v^+)$, the loss treats a (potential) positive as negative, pushing $u$ away from $v$ and inducing counterproductive updates.
(ii) \emph{Non-zero bias:} whenever $\pi(u,v^+)>0$ and $\phi'(\cdot)\neq 0$, the expected gradient is biased.
(iii) \emph{Bias scales with FHN mass:} letting
$\Delta(u,v^+)=\mathbb{E}_{q_{\mathrm{FHN}}}[\nabla_\theta \ell]-\mathbb{E}_{q_{\mathrm{TN}}}[\nabla_\theta \ell]$,
we have $\|\mathrm{Bias}(u,v^+)\|\approx \pi(u,v^+)\|\Delta(u,v^+)\|$. Under a mild non-degeneracy condition $\|\Delta(u,v^+)\|\ge C$ over a non-trivial training region, $\|\mathrm{Bias}(u,v^+)\|\ge C\cdot \pi(u,v^+)$.

\paragraph{4) HNLMRec: Bridging the gap via conditional probability shift.}
HNLMRec alters the naive semantic sampling distribution by injecting collaborative signals into the LLM conditioning through hybrid prompts and alignment objectives.
We denote by $q_\theta(\cdot \mid \cdot)$ the sampling distribution induced by the LLM under a fixed decoding rule $\mathcal{D}$ (e.g., temperature/top-$k$/top-$p$).
Accordingly, we define
\begin{equation}
\begin{aligned}
& q_{\text{aligned}}(v \mid u, v^+)  \;=\; \\ &
q_\theta\!\left(v \mid \mathcal{N}_S(v^+),\, \psi(\mathbf{e}_u),\, \psi(\mathbf{e}_{v^+}),\, \mathcal{P}_u,\, \mathcal{P}_{v^+}\right).
\end{aligned}
\label{eq: q_alig}
\end{equation}
Here $\mathbf{e}_u,\mathbf{e}_{v^+}$ are collaborative embeddings from a CF model, $\psi(\cdot)$ maps them to the LLM representation space, and $\mathcal{P}_u,\mathcal{P}_{v^+}$ are semantic profiles.
Since the SFT target is a mined collaborative negative (i.e., structurally distant from $\mathbf{e}_u$), optimizing the SFT objective updates $\theta$ and imposes a \emph{soft behavioral constraint} in the LLM representations, shifting probability mass away from the FHN region $\mathcal{N}_S(v^+)\cap \mathcal{I}_u^+$ and toward the true-negative region $\mathcal{N}_S(v^+)\setminus \mathcal{I}_u^+$.
Equivalently, HNLMRec reduces the aligned noise rate:
\begin{equation}
\begin{aligned}
\pi_{\text{aligned}}(u,v^+) & =\; \mathbb{P}_{v\sim q_{\text{aligned}}(\cdot\mid u,v^+)}\!\left[v\in \mathcal{I}_u^+\right],
\\ &
\pi_{\text{aligned}}(u,v^+) \ll \pi(u,v^+).
\label{eq:pi_aligned}
\end{aligned}
\end{equation}

\paragraph{5) Information-theoretic view: manifold alignment via Mutual Information maximization.}
The core mechanism driving the shift is the contrastive SFT / alignment objective.
From an information-theoretic perspective, minimizing an InfoNCE-style loss maximizes a lower bound of the mutual information (MI) between the LLM representation and the collaborative negative label:
\begin{equation}
\min \mathcal{L}_{\mathrm{InfoNCE}}
\quad \Longleftrightarrow \quad
\max \; I(\mathbf{Q}_{\mathrm{emb}};\mathbf{N}^{+}_{\mathrm{emb}}),
\label{eq: mi_view}
\end{equation}
where $\mathbf{Q}_{\mathrm{emb}}$ is the LLM query embedding induced by the hybrid prompt and $\mathbf{N}^{+}_{\mathrm{emb}}$ is the collaborative hard negative label embedding.
Maximizing this MI achieves two effects:
(i) \emph{Hallucination filtration:} $\mathbf{N}^{+}_{\mathrm{emb}}$ lies on the valid collaborative item manifold $\mathcal{M}_c$; a hallucinated output behaves like noise and yields near-zero MI, thus MI maximization encourages the LLM to produce manifold-consistent representations.
(ii) \emph{Semantic--behavioral orthogonalization:} collaborative negatives are chosen to be behaviorally distinct from user history (far from $\mathbf{e}_u$); aligning $\mathbf{Q}_{\mathrm{emb}}$ to such negatives implicitly pushes the generated representation away from the user's positive region.

\paragraph{6) Bias cancellation.}
Combining the conditional probability shift in Eq.~\eqref{eq: q_alig} and the MI-driven alignment in Eq.~\eqref{eq: mi_view}, HNLMRec constructs $q_{\text{aligned}}$ that approximates the ideal $q^\star$ by suppressing the FHN mass.
Replacing $\pi(u,v^+)$ in Eq.~\eqref{eq:grad_bias_decomp} with $\pi_{\text{aligned}}(u,v^+)$ yields
\begin{equation}
\begin{aligned}
\|\mathrm{Bias}_{\text{aligned}}(u,v^+)\|
 & \approx 
\pi_{\text{aligned}}(u,v^+) \\ &\cdot \|\Delta_{\text{aligned}}(u,v^+)\|.
\end{aligned}
\end{equation}
Therefore, as $\pi_{\text{aligned}}(u,v^+)$ decreases, the gradient bias is asymptotically eliminated:
\begin{equation}
\begin{aligned}
\pi_{\text{aligned}}(u,v^+) & \to  0 
\quad   \\ &\Longrightarrow \quad
\|\mathrm{Bias}_{\text{aligned}}(u,v^+)\|\to 0.
\end{aligned}
\end{equation}
\textbf{This explains why HNLMRec can exploit semantic hardness while avoiding behavioral noise, enabling more stable and unbiased optimization.} The analysis of the model's time complexity is presented in the Appendix~\ref{ap: times_analy}. The pseudocode of the HNLMRec is provided in the Algorithm~\ref{alg: llm-driven-sampling}. We present related work in the Appendix~\ref{ap: relateed work}.

\subsection{Discussion}
In this section, we clarify how LLM-based data synthesis fundamentally differs from self-supervised learning (SSL) augmentations. While SSL-based recommenders demonstrate the effectiveness of augmentation, our approach departs from them in three key aspects:
\textbf{(1) Open-world knowledge.} Graph perturbations (edge dropout/masking) are closed-world and cannot introduce new semantics, whereas HNLMRec leverages LLM knowledge to synthesize virtual items beyond observed structures.
\textbf{(2) Grounded counterfactuals.} Random noise lacks semantic control and may destroy preference signals; HNLMRec aligns generation with collaborative signals to produce semantically plausible but behaviorally irrelevant hard negatives.
\textbf{(3) Manifold bridging.} Conventional augmentations operate only on the collaborative manifold, while HNLMRec explicitly aligns semantic and collaborative manifolds, reducing false negatives from naive semantic augmentation.

\begin{table*}[htbp]
\centering
\caption{
The table below compares the performance of various competing methods and HNLMRec across two datasets. Bold text indicates the highest score, while underlining denotes the best result among the baseline methods. An asterisk (*) signifies a statistically significant improvement over the best baseline method (p < 0.05).
}
\vspace{-0.3cm}
\setlength{\tabcolsep}{3pt}
\renewcommand{\arraystretch}{1}
\label{tab:comparison}
\begin{tabular}{cc|cccc|cccc}
\hline
\multicolumn{2}{c|}{\multirow{2}{*}{\textbf{Backbone Models}}}&
\multicolumn{4}{c|}{\textbf{Toys \& Games}} &
\multicolumn{4}{c}{\textbf{CDs \& Vinyl}} \\
\cline{3-10}
 & & \textbf{R@10} & \textbf{N@10} & \textbf{R@20} & \textbf{N@20}
   & \textbf{R@10} & \textbf{N@10} & \textbf{R@20} & \textbf{N@20}  \\
\hline 
\multirow{9}{*}{\textbf{MF}} & -RNS & 1.48 & 0.93 & 2.57 & 1.14 & 2.01 & 1.53 & 4.47 & 2.37 \\
 & -DNS & 1.81 & 1.18 & 2.84 & 1.52 & 2.91 & 2.03 & 4.63 & 2.55 \\
 & -MixGCF & 1.96 & 1.22 & 2.72 & 1.51 & 3.00 & 2.15 & 4.96 & 2.64 \\
 & -AHNS & 1.84 & 1.23 & 3.11 & 1.63 & 2.84 & 1.79 & 4.49 & 2.59 \\
 & -KAR & 1.83 & 1.22 & 3.18 & 1.63 & 2.83 & 1.78 & 4.48 & 2.58 \\
 & -LLMRec & 1.87 & 1.28 & 2.73 & 1.52 & 2.84 & 1.79 & 4.49 & 2.59 \\
 & -RLMRec & 2.04 & 1.23 & 3.19 & 1.64 & 2.92 & 2.04 & 4.97 & 2.67 \\
 & -AlphaRec & \underline{2.15} & \underline{1.35} & \underline{3.35} & \underline{1.72}
             & \underline{3.10} & \underline{2.22} & \underline{5.05} & \underline{2.75} \\
 \rowcolor{blue!15} & \textbf{OURS} 
   & \textbf{2.33*} & \textbf{1.45*} & \textbf{3.51*} & \textbf{1.82*}
   & \textbf{3.24*} & \textbf{2.33*} & \textbf{5.21*} & \textbf{2.90*} \\
\hline
\multirow{9}{*}{\textbf{LightGCN}} & -RNS & 2.05 & 1.35 & 3.37 & 1.78 & 3.09 & 1.75 & 5.06 & 2.53 \\
 & -DNS & 2.12 & 1.39 & 3.41 & 1.78 & 3.19 & 2.10 & 5.43 & 2.71 \\
 & -MixGCF & 2.27 & 1.43 & 3.52 & 1.83 & 3.34 & 2.11 & 5.65 & 2.81 \\
 & -AHNS & 2.21 & 1.38 & 3.66 & 1.82 & 3.28 & 2.00 & 5.59 & 2.75 \\
 & -KAR & 2.20 & 1.36 & 3.65 & 1.81 & 3.27 & 1.99 & 5.44 & 2.74 \\
 & -LLMRec & 2.20 & 1.37 & 3.64 & 1.80 & 3.26 & 1.98 & 5.69 & 2.73 \\
 & -RLMRec & 2.25 & 1.40 & 3.67 & 1.84 & 3.25 & 2.09 & 5.77 & 2.85 \\
 & -AlphaRec & \underline{2.33} & \underline{1.52} & \underline{3.78} & \underline{1.92}
             & \underline{3.40} & \underline{2.15} & \underline{5.85} & \underline{2.93} \\
 \rowcolor{blue!15} &  \textbf{OURS} 
   & \textbf{2.40*} & \textbf{1.64*} & \textbf{3.92*} & \textbf{1.99*}
   & \textbf{3.53*} & \textbf{2.32*} & \textbf{5.99*} & \textbf{3.11*} \\
\hline
\end{tabular}
\vspace{-0.5cm}
\end{table*}

\section{Experiments}
\subsection{Experimental Settings}
\textbf{Datasets.}
We conduct experiments on four real-world datasets: Toys \& Games, CDs \& Vinyl, Yelp2018, and Amazon Electronics. The first three datasets are used for fine-tuning, while Amazon serves as a held-out benchmark to evaluate the generalization performance of HNLMRec. Dataset statistics are reported in Table~\ref{tab:dataset_stats}. The detailed dataset preprocessing procedures are provided in the Appendix~\ref{ap: data_process}.
\textbf{Baselines}. We use the following models as baselines: MF~\cite{koren2009matrix}, 
NGCF~\cite{wang2019neural}, LightGCN~\cite{he2020lightgcn}, RNS~\cite{koren2009matrix}, DNS~\cite{shi2023theories}, MixGCF~\cite{huang2021mixgcf}, AHNS~\cite{lai2024adaptive}, KAR~\cite{xi2024towards}, RLMRec~\cite{ren2024representation}, LLMRec~\cite{wei2024llmrec}, AlphaRec~\cite{sheng2024language}. Model descriptions and implementation details are provided in Appendices~\ref{ap: baselines} and~\ref{ap: details}.

\subsection{Overall Performance}
Table~\ref{tab:comparison} summarizes the results on Toys \& Games and CDs \& Vinyl. HNLMRec achieves the best performance across \emph{all} metrics on both datasets, with statistically significant improvements over the strongest baseline (AlphaRec). Concretely, under the MF backbone, HNLMRec increases NDCG@20 on Toys from 1.72 to 1.82, yielding a 5.81\% gain, and improves NDCG@20 on CDs from 2.75 to 2.90 (a 5.45\% gain). Similar gains persist with stronger backbones: under LightGCN, NDCG@10 improves from 1.52 to 1.64 on Toys (7.89\%) and from 2.15 to 2.32 on CDs (7.91\%), accompanied by consistent Recall improvements. The superiority of HNLMRec is also model-agnostic and extends to NGCF (see Table~\ref{tab: comparison_main} in Appendix); for example, Toys NDCG@20 rises from 1.62 to 1.72 (6.17\%). We observe the same trend on Yelp2018 (see Table~\ref{tab:yelp_comparison} in Appendix), where HNLMRec yields stable and significant gains across different backbones. Finally, other baselines (such as LLMRec and RLMRec) are generally less competitive than strong ID-based hard negative samplers, whereas HNLMRec consistently outperforms them, supporting our claim that collaborative-semantic alignment suppresses FHNS and corrects the optimization direction, thereby validating the effectiveness of leveraging LLMs to generate negatives.

\begin{table*}[!t]
\centering
\renewcommand{\arraystretch}{0.95}
\setlength{\tabcolsep}{3pt}
\caption{Ablation results on Toys \& Games and CDs \& Vinyl. The best results are in \textbf{bold}, and the best baseline results are \underline{underlined}.  $^{*}$ indicates a statistically significant improvement ($p<0.05$, t-test) over the best baseline.}
\vspace{-0.3cm}
\label{tab: empirical-toys-cds}
\begin{tabular}{lcccccccc}
\toprule
\multirow{2}{*}{\textbf{Method}} & \multicolumn{4}{c}{\textbf{Toys \& Games}} & \multicolumn{4}{c}{\textbf{CDs \& Vinyl}} \\
\cmidrule(lr){2-5}\cmidrule(lr){6-9}
& \textbf{R@10} & \textbf{N@10} & \textbf{R@20} & \textbf{N@20} & \textbf{R@10} & \textbf{N@10} & \textbf{R@20} & \textbf{N@20} \\
\midrule
MF       & 1.58 & 1.03 & 2.67 & 1.34 & 2.62 & 1.83 & 4.47 & 2.37 \\
NGCF     & 1.56 & 0.92 & 3.16 & 1.40 & 2.65 & 1.60 & 5.00 & 2.30 \\
LightGCN & 2.05 & 1.35 & 3.46 & 1.78 & 3.09 & 1.95 & 5.65 & 2.73 \\
\midrule
\multicolumn{9}{c}{\textbf{LLMs w/o Alignment}} \\
\addlinespace[2pt]
\textit{Qwen2-0.5B-Instruct}   & 1.98 & 1.25 & 3.22 & 1.60 & 2.92 & 1.81 & 5.39 & 2.56 \\
\textit{Qwen2.5-7B-Instruct}   & 2.07 & 1.43 & 3.63 & 1.86 & 3.23 & 2.01 & 5.80 & 2.82 \\
\textit{Llama3-8B-Instruct}    & 2.20 & 1.41 & 3.65 & 1.86 & 3.30 & 2.08 & 5.82 & 2.84 \\
\textit{Qwen2.5-72B-Instruct}  & 2.24 & 1.48 & 3.70 & 1.87 & 3.33 & 2.11 & 5.86 & 2.85 \\
\textit{ChatGPT-3.5}           & 2.25 & 1.58 & 3.73 & \underline{1.90} & 3.37 & 2.12 & \underline{5.90} & \underline{2.88} \\
\midrule
\multicolumn{9}{c}{\textbf{LLMs w/ Alignment}} \\
\addlinespace[2pt]
\textit{Qwen2.5-7B-Instruct}   & \underline{2.38} & \underline{1.63} & \underline{3.91} & 1.89 & \underline{3.43} & \underline{2.21} & 5.84 & 2.82 \\
\rowcolor{blue!15} \textit{Llama3-8B-Instruct}    & \textbf{2.40*} & \textbf{1.64*} & \textbf{3.92*} & \textbf{1.99*} & \textbf{3.52*} & \textbf{2.22*} & \textbf{5.99*} & \textbf{3.01*} \\
\bottomrule
\end{tabular}
\vspace{-0.5cm}
\end{table*}

\subsection{Ablation Study}
We conduct ablation studies with LightGCN as the backbone, and report the results in Table~\ref{tab: empirical-toys-cds}. Here, \textit{LLMs w/o Alignment} denotes directly using LLM-generated semantics without imposing collaborative alignment constraints, while \textit{LLMs w/ Alignment} injects collaborative signals and applies a contrastive alignment objective to explicitly map semantic representations onto the collaborative space, thereby mitigating the \emph{Semantic--Behavior Gap}. From Table~\ref{tab: empirical-toys-cds}, we obtain three key observations: (1) \textbf{Alignment is crucial}: compared with \textit{w/o Alignment}, introducing alignment yields consistent improvements across all metrics on both datasets, indicating that without alignment, semantic generation is prone to false hard negatives and weakened supervision. (2) \textbf{Model scale is not the decisive factor}: without alignment, increasing model size brings limited and unstable gains; with alignment, a medium-sized model can achieve stable and superior performance, suggesting that structured collaborative--semantic constraints matter more than stronger semantic priors alone. (3) \textbf{Alignment improves stability and transferability}: results under \textit{w/o Alignment} vary noticeably across LLMs, whereas \textit{w/ Alignment} delivers more consistent advantages across datasets and metrics, validating that collaborative--semantic alignment filters false hard negatives and corrects the optimization direction, thus improving the effectiveness of LLM-synthesized negatives. 

\vspace{-0.33cm}
\subsection{Hyper-parameter Analysis}
Figure~\ref{fig: hyper_paramers} presents the hyperparameter sensitivity and convergence analyses (NDCG@20). In (a), performance on Toys and CDs remains largely stable as the number of negatives varies, with only slight gains; however, using more negatives increases training time and computational cost, indicating that the method is relatively insensitive to this hyperparameter and robust in practice. In (b)(c), HNLMRec converges faster and achieves a higher final performance during training, demonstrating superior optimization efficiency and stability compared to Mixup and DNS.

\subsection{Potentials of HNLMRec}
\textbf{Robustness Under Sparse Data}.
When interactions are scarce, ID-based negative sampling tends to produce “easy” negatives, resulting in weak supervision. Figure \ref{fig: combined_analysis}(a) shows that HNLMRec consistently outperforms the baselines across different training-data ratios, benefiting from LLMs’ semantic understanding that can still capture user preferences and mine high-quality hard negatives under limited interactions, thereby alleviating sparsity.
\textbf{High Performance Under Popularity Bias}.
Following \cite{yu2022graph}, we split the test set by popularity into Unpopular/Normal/Popular. As shown in Figure \ref{fig: combined_analysis}(b), HNLMRec gains the most on Unpopular items, as LLM semantics enhance preference modeling, whereas LightGCN with random negatives performs better on Popular items, reflecting its popularity bias.  \textbf{Due to space constraints}, we defer additional analyses, results, and discussions on generalization to Appendix~\ref{ap: potentials}.

\section{Conclusion}
This work proposes HNLMRec, which  mitigates the FHNS issue in conventional hard negative sampling by \emph{directly synthesizing} negative samples. HNLMRec aligns the semantic space with collaborative signals via contrastive fine-tuning, generating negatives that are \emph{semantically hard yet behaviorally irrelevant}, and performs synthesis in the embedding space to improve efficiency and compatibility. Extensive experiments across multiple datasets demonstrate that HNLMRec significantly outperforms existing baselines.

\section{Limitations}
Although HNLMRec can consistently improve the quality of hard negatives, it introduces additional training overhead due to the contrastive fine-tuning and the synthesis module, which may require more careful compute budgeting in large-scale deployments. Moreover, while embedding-space synthesis improves efficiency and compatibility, its effectiveness can still depend on the quality of item text/profile signals and the underlying CF backbone. Finally, our evaluation mainly focuses on offline benchmarks, and further validating HNLMRec in fully online settings with real-time feedback remains an interesting direction for future work.

\bibliography{custom}
\appendix
\clearpage
\section{Related Work}
\label{ap: relateed work}
\subsection{Negative Sampling in Recommendation}
According to \cite{wu2023dimension}, negative sampling methods in recommender systems are classified into Point-wise and Line-wise Sampling Methods. \textbf{Point-wise sampling} directly selects negative samples from the candidate set, with static methods often randomly choosing items not interacted with by users \cite{he2017neural,he2023simplifying,wang2019kgat,rendle2012bpr, yuan2023knowledge}. Popular-based methods tend to select more frequent items as negatives \cite{cheng2021learning,wang2019multi}. However, these methods lack adaptability and result in lower quality negatives. Dynamic methods have been proposed to adjust sampling strategies based on training status or user behavior \cite{ying2018graph,gong2022itsm,shi2023theories}, but they still depend on the candidate pool, limiting the extraction of high-quality negatives.
In contrast, \textbf{Line-wise sampling} methods improve effectiveness by generating pseudo-negatives. For instance, MixGCF \cite{huang2021mixgcf} interpolates between positive and negative samples, while DINS \cite{wu2023dimension} employs boundary definitions and multi-hop pooling for flexibility. However, point sampling quality is affected by the candidate pool and data sparsity, and line sampling often focuses on popular samples, missing long-tail item characteristics. This impacts recommendations for cold-start users. Hence, this paper proposes employing LLM to accurately characterize user preferences and address these challenges through semantic negative sampling.
\subsection{LLM-based Recommendation}
 Existing LLM-based recommendation algorithms can be divided into two categories \cite{wu2024survey,li2023large,zhao2023recommender}: using LLMs as recommenders and using LLMs to enhance traditional recommendations. The first method primarily recommends items through text generation paradigms. In earlier studies, researchers mainly focused on designing specific prompts or contextual learning strategies to adapt LLMs to downstream recommendation tasks \cite{li2023prompt, tachioka2024user, liao2024llara,acharya2023llm}. However, these methods often perform worse than traditional models. With the development of LLM fine-tuning techniques, recent work has primarily focused on fine-tuning LLMs using recommendation-related corpora and aligning LLMs with traditional recommendations \cite{lin2024data, li2023guided,bao2023tallrec}. For instance, coLLM \cite{zhang2023collm}
 maps collaborative information into the latent space of LLMs for representation alignment to improve recommendation performance. The second method enhances traditional recommender systems by utilizing semantic representations generated by LLMs \cite{sun2024large, zheng2024adapting, zhu2024collaborative}. 
 This process typically involves leveraging LLMs to analyze user and item attributes, construct profiles, generate embeddings, and integrate them into traditional recommender systems. In this work, we explore using LLMs for semantic negative sampling to mine hard negative samples from user-item pairs, aiming to enhance the performance of traditional recommendation models.

 \paragraph{LLM-assisted sample selection and negative generation.}
Recent works have explored leveraging (multi-modal) LLMs to improve training signals for recommendation, yet their goals and mechanisms differ from ours.
LLMHD~\cite{song2024large} targets \emph{denoising recommendation} by using an LLM as a \emph{scorer} to distinguish hard samples from noisy interactions via semantic-consistency scoring, coupled with variance-based pruning and iterative preference updating to reduce the cost of LLM calls and refine user preference summaries.
In contrast, our method uses LLMs as a \emph{generator} to synthesize semantically plausible hard negatives and explicitly enforces behavioral validity by aligning the generated hard-negative representation with a collaborative hard-negative label mined by the CF backbone.
Meanwhile, NegGen~\cite{ji2025generating} studies \emph{multi-modal} recommendation and generates balanced, contrastive negatives by manipulating item attributes with MLLM prompts and further introduces a causal module to disentangle intervened key features from irrelevant attributes.
Different from NegGen, we focus on the semantic--behavior mismatch in implicit CF that induces false hard negatives, and propose a collaborative-aligned generation-and-alignment scheme to mitigate such FHNS-driven supervision noise.

\section{Time Complexity Analysis}
\label{ap: times_analy}
The overall workflow of HNLMRec consists of two main stages: (1) a supervised fine-tuning stage, where the LLM is optimized using collaborative signals to generate high-quality hard negative embeddings; and (2) an inference-and-training stage, where the fine-tuned LLM is used to pre-generate hard negative embeddings for all training user-item pairs before training the collaborative filtering model (e.g., LightGCN). 
Notably, instead of invoking the LLM during each training step, which would be computationally expensive. HNLMRec adopts an offline preprocessing strategy, where hard negative embeddings are generated and cached in advance. During training, only constant-time lookup operations are required, thus ensuring high efficiency. This design shifts the computational burden of the LLM to the offline stage, making the actual training process comparable to that of the base CF model, with minimal additional overhead.
Specifically, HNLMRec introduces three additional operations during training: 
\textbf{(1)} a dimensionality reduction module that maps high-dimensional LLM embeddings to the CF latent space; 
\textbf{(2)} a contrastive alignment module that aligns user, positive, and hard negative embeddings; and \textbf{(3)} an embedding lookup operation for retrieving hard negatives. To evaluate the overall efficiency, we analyze the time complexity of HNLMRec using LightGCN as the backbone.

We analyze the time complexity of HNLMRec using the following notations: let $N$ denote the number of training interactions, $d_{\text{LLM}}$ the dimensionality of the high-dimensional embeddings generated by the LLM, $d$ the embedding dimension used in the recommendation model, $L$ the number of GNN propagation layers, $B$ the batch size, and $E$ the number of training epochs. In the offline stage, HNLMRec maps each high-dimensional embedding generated by the LLM to the recommendation embedding space through a two-layer MLP, with a total complexity of $\mathcal{O}(N \cdot d_{\text{LLM}} \cdot d)$. During training, the cost of looking up pre-generated hard negative embeddings is constant at $\mathcal{O}(1)$, and the BPR score computation per batch requires $\mathcal{O}(B \cdot d)$. The graph propagation and aggregation process of LightGCN has a complexity of $\mathcal{O}(L \cdot (M + V + N) \cdot d)$, where $M$ and $V$ are the numbers of users and items, respectively. In addition, the contrastive alignment module computes cosine similarities between each anchor user and its corresponding positive and negative samples, resulting in a complexity of $\mathcal{O}(B \cdot d)$. Therefore, compared to the original LightGCN, HNLMRec introduces only a constant overhead from embedding fusion and contrastive learning. \textbf{The overall training complexity remains within an acceptable range, demonstrating good scalability for large-scale recommendation scenarios}.

\begin{figure*}[t]
    \centering
    \begin{minipage}{0.25\linewidth}
        \centerline{\includegraphics[width=\textwidth]{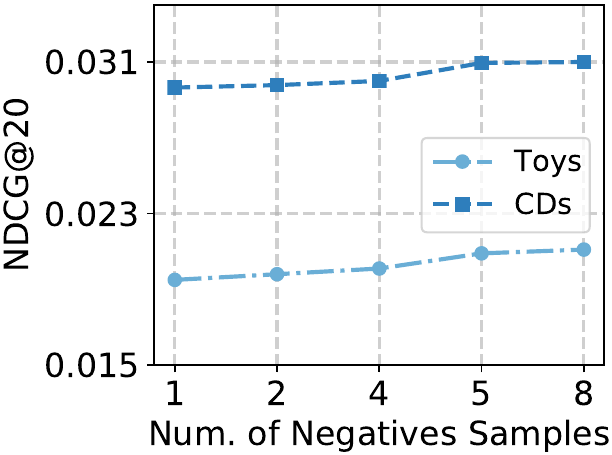}}
        \centerline{\;\;\;\;\;\;\;\;\;\;(a)}
    \end{minipage}
    \begin{minipage}{0.25\linewidth}
    \centerline{\includegraphics[width=\textwidth]{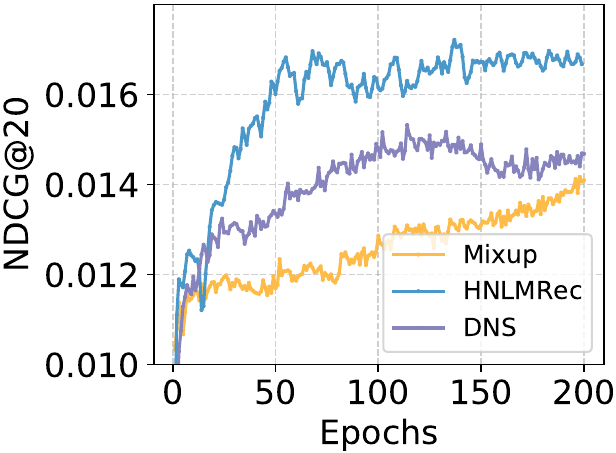}}
        \centerline{\;\;\;\;\;\;\;\;\;\;(b)}
    \end{minipage}
    \begin{minipage}{0.25\linewidth}
    \centerline{\includegraphics[width=\textwidth]{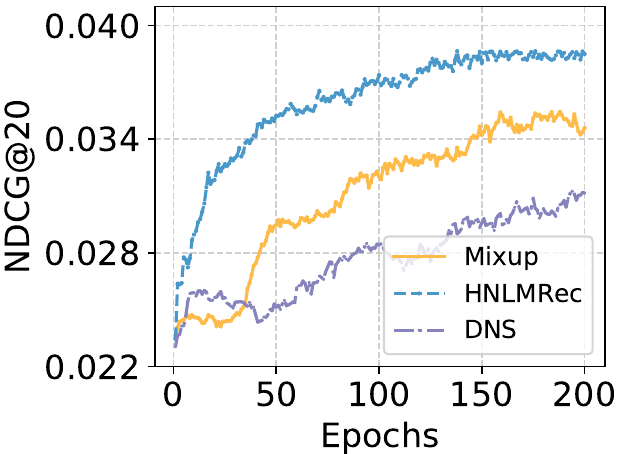}}
        \centerline{\;\;\;\;\;\;\;\;\;\;(c)}
    \end{minipage}
    \caption{Figures (a) analyze the effects of the negative sample size. Figures (b) and (c) compare the convergence speed of HNLMRec against ID-based negative sampling baselines on Toys and Yelp.}
    \label{fig: hyper_paramers}
\end{figure*}

\begin{table}[!t]
\centering
\caption{Dataset Statistics}
\label{tab:dataset_stats}
\setlength{\tabcolsep}{3pt}
\renewcommand{\arraystretch}{1.15}
\begin{tabular}{cccc}
\hline
\textbf{Dataset} & \textbf{\#Users} & \textbf{\#Items} & \textbf{\#Interactions} \\ \hline
Toys \& Games & 22,338 & 9,023  & 200,511 \\
CDs \& Vinyl & 19,385 & 8,279 & 186,535 \\
Yelp2018 & 29,832 & 16,781 & 513,976 \\
Amazon & 97,570  & 44,669 & 178,259 \\
\hline
\end{tabular}
\end{table}
\section{Baselines and Datasets}
\subsection{Dataset Processing Details}
\label{ap: data_process}
The preprocessing steps for each dataset are as follows:
For the Toys \& Games dataset, we first filtered out interactions with ratings lower than 4 and retained records within the date range from 2015-01-01 to 2018-01-01. We further removed users and items with fewer than 10 interactions.
For the CDs \& Vinyl dataset, we selected records from 2014-01-01 to 2016-01-01, while applying the same filtering criteria as above.
For the Yelp2018 dataset, we retained records within the range of 2015-01-01 to 2018-01-01, with the same filtering strategy as applied to the above datasets.
For the Amazon Electronics  dataset, we preserved only interactions with ratings of 3 or higher and selected records within the date range from 2023-01-01 to 2023-09-30.

\subsection{Baselines}
\label{ap: baselines}
We categorize the baselines into three major groups based on their modeling paradigms and enhancement strategies:

\noindent\textbf{Traditional Collaborative Filtering Models}:

\begin{itemize}[leftmargin=0pt]
   \item \textbf{MF} \cite{koren2009matrix}: Matrix Factorization is a classical latent factor model that represents users and items in a shared embedding space and estimates preferences through inner product.
    \item \textbf{NGCF} \cite{wang2019neural}: Neural Graph Collaborative Filtering extends MF by modeling high-order user-item connectivity on interaction graphs via graph neural networks (GNNs), enabling better exploitation of collaborative signals.
    \item \textbf{LightGCN} \cite{he2020lightgcn}: A streamlined GNN-based CF model that removes unnecessary nonlinearities and feature transformations to retain only essential neighborhood aggregation, significantly improving efficiency and effectiveness.
     
  \textbf{Negative Sampling Methods}:
  
         \item \textbf{RNS} \cite{koren2009matrix}: Random Negative Sampling is a widely adopted baseline that randomly samples non-interacted items as negatives, though it often produces overly easy samples.
        \item \textbf{DNS} \cite{shi2023theories}: Dynamic Negative Sampling adaptively selects harder negatives from a sampled pool by comparing model scores, offering stronger training signals and faster convergence.
        \item \textbf{MixGCF} \cite{huang2021mixgcf}: This method augments the negative sample set by injecting positive samples into the negative pool, thereby synthesizing more semantically challenging negatives and improving model discrimination.
        \item \textbf{AHNS} \cite{lai2024adaptive}: Adaptive Hard Negative Sampling dynamically measures sample difficulty during training and selects informative negatives based on learned item similarities, balancing hardness and diversity.
        
       \noindent\textbf{LLM-Enhanced Recommendation Models}:
        \item\textbf{KAR} \cite{xi2024towards}: Knowledge-Aware Recommendation leverages LLMs to generate detailed user and item descriptions from side information (e.g., reviews), which are then fused into traditional recommendation models to enhance semantic understanding.
        \item\textbf{RLMRec} \cite{ren2024representation}: This model constructs richer user/item embeddings by integrating LLM-generated semantic texts, improving generalization and robustness of downstream CF models.
       \item\textbf{LLMRec} \cite{wei2024llmrec}: A recent method that enhances ID-based models with attribute-level knowledge produced by LLMs, facilitating improved personalization through richer context modeling and attribute reasoning. 
       \item \textbf{AlphaRec}~\cite{sheng2024language} is a recommender that directly uses pretrained language-model (LM) text representations as the backbone for collaborative filtering (CF). The paper shows that simply projecting item-text embeddings (e.g., titles) into the recommendation space with a linear map already yields strong item representations and competitive performance, suggesting a structural correspondence between LM and CF embedding spaces—i.e., LMs may implicitly encode collaborative signals.

In the ablation study (Table~\ref{tab: empirical-toys-cds}), we further evaluate the effect of LLM scale by comparing multiple backbone sizes, including Qwen2-0.5B-Instruct~\cite{team2024qwen2}, Qwen2.5-7B-Instruct~\cite{yang2025qwen3}, Llama3-8B-Instruct~\cite{dubey2024llama}, Qwen2.5-72B-Instruct~\cite{yang2025qwen3}, and ChatGPT-3.5~\cite{brown2020language}.

\end{itemize}
\subsection{Implementation Details}
\label{ap: details}
We chose \textit{Llama3-8B-Instruct} as the base model in the model fine-tuning task. We collect 100,000 data points from Yelp2018, Toys, and CDs and constructed the dataset required for fine-tuning according to Equation (\ref{eq: prompt_const}). The fine-tuning process is done on a system equipped with eight NVIDIA 4090 GPUs, utilizing 4-bit quantization technology. To ensure a fair comparison, all LLMs in the baseline used Llama3-8B as the base model. 
For all GNN-based methods, we fix the number of propagation layers to 3 and the embedding dimension to 64, and tune the learning rate from $\{1e-4, 1e-3, 2e-3, 1e-2\}$.  For candidate-pool-based negative sampling methods (DNS, MixGCF, and AHNS), we set the number of selected negatives per positive interaction to 2. For \textsc{HNLMRec}, we also use 2 negatives per positive, and defer the detailed processing strategy of multiple synthesized negatives to Section~\ref{ap: neg_process}.

\section{ How to process Multiple LLM-generated Negatives}
\label{ap: neg_process}
For each observed positive interaction $(u, v^+)$, the LLM synthesizes a candidate set of $K$ negatives:
\begin{equation}
\mathcal{N}(u,v^+) = \{v^-_1,\dots,v^-_K\}.
\end{equation}
Instead of using all candidates, we select the most informative Top-1 (or Top-$k$) negatives for training to reduce noise accumulation and strengthen the learning signal. Specifically, we adopt a model-aware hardness criterion and pick the negatives that receive the highest scores under the current model:
\begin{equation}
v^- = \arg\max_{v \in \mathcal{N}(u,v^+)} f_\theta(u,v),
\end{equation}
where $f_\theta(u,v)$ denotes the model scoring function. Intuitively, this strategy prioritizes the negatives closest to the current decision boundary, yielding harder supervision and more effective gradients. For Top-$k$ selection, we choose the $k$ highest-scoring items from $\mathcal{N}(u,v^+)$ according to $f_\theta(u,v)$, and aggregate their losses (e.g., by averaging) during optimization.

\section{Potentials of LLM-Driven Hard Negative Sampling}
\label{ap: potentials}
\begin{figure}[t]
    \centering
    \begin{minipage}{0.48\linewidth}
        \centering
        \includegraphics[width=\textwidth]{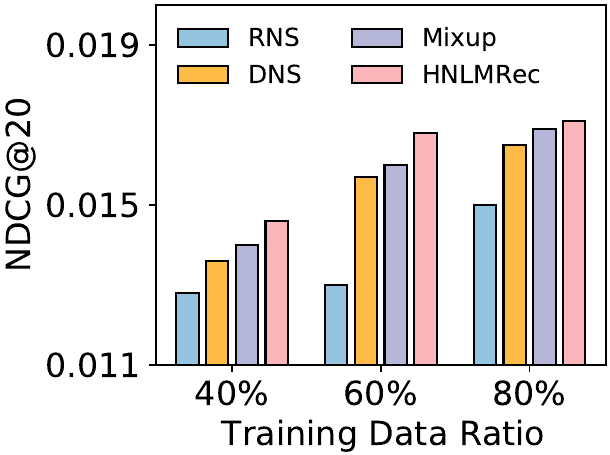}
        \centerline{\small (a) Data Sparsity (Toys)}
    \end{minipage}
    \begin{minipage}{0.48\linewidth}
        \centering
        \includegraphics[width=\textwidth]{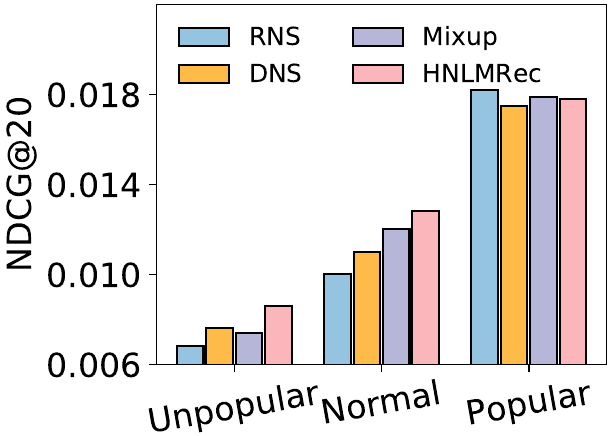}
        \centerline{\small (b) Popularity Groups}
    \end{minipage}
    \vspace{-0.3cm}
    \caption{Model potential: (a) sparsity study on Toys with varying training data ratios; (b) performance across item popularity groups.}
    \label{fig: combined_analysis}
    \vspace{-0.5cm}
\end{figure}

\textbf{Good Generalization on New Dataset}. 
We test the performance of HNLMRec, fine-tuned on other datasets, against ID-based negative sampling methods and LLM-based methods using LightGCN as the backbone model on the Amazon Fashion dataset. As shown in Table \ref{tab:gen}, HNLMRec significantly outperforms other baseline models, demonstrating its strong generalization capability on new datasets without requiring additional fine-tuning for specific datasets. We attribute this to the powerful generalization ability of LLMs, the domain adaptability gained through fine-tuning, the robustness of semantic negative sampling, and the common characteristics across datasets. These factors collectively enable HNLMRec to maintain excellent performance across different datasets. 

\textbf{Robust Performance Under Sparse Data Conditions}.
Data sparsity is one of the main challenges faced by ID-based negative sampling methods. Due to the limited historical interaction data of users, traditional negative sampling methods often generate negative samples that lack challenge, making it difficult to effectively distinguish between items that users truly like and those they are not interested in, thereby affecting recommendation accuracy. In Figure \ref{fig: combined_analysis} (a), we compare the performance of HNLMRec with baseline models under different proportions of training data. The experimental results show that HNLMRec significantly outperforms the baseline models. We attribute this advantage to the powerful semantic understanding capability of LLMs, which can accurately capture user preferences even with limited interaction data, thereby enabling high-quality hard negative sample mining and improving model performance. The experimental results further demonstrate the potential of LLMs in alleviating data sparsity issues through precise hard negative sample sampling.

\begin{table}[t!]
\centering
\caption{Overall performance comparison on the Amazon Electronics  dataset}\vspace{-0.3cm}
\setlength\tabcolsep{3.2pt}
\begin{tabular}{l|cccc}
\toprule
\textbf{Methods} & \textbf{R@10} & \textbf{R@20} & \textbf{N@10} & \textbf{N@20} \\
\midrule
RNS     & 0.76 & 0.48 & 1.19 & 0.58 \\
DNS     & 0.85 & 0.50 & 1.22 & 0.60 \\
MixGCF  & \underline{0.97} & 0.52 & 1.27 & \underline{0.62} \\
AHNS    & 0.90 & 0.51 & 1.25 & 0.61 \\
\midrule
KAR     & 0.78 & 0.49 & 1.20 & 0.59 \\
LLMRec  & 0.91 & \underline{0.58} & 1.26 & 0.60 \\
RLMRec  & 0.96 & 0.54 & \underline{1.30} & \underline{0.62} \\
\midrule
\rowcolor{blue!10} HNLMRec & \textbf{0.98$^{*}$} & \textbf{0.60$^{*}$} & \textbf{1.32$^{*}$} & \textbf{0.64$^{*}$} \\
\bottomrule
\end{tabular}
\vspace{-0.3cm}
\label{tab:gen}
\end{table}

\textbf{Achieving High Performance on Popularity Bias Distributions}.
Popular items dominate interactions in RS due to frequent recommendations, while long-tail item interactions are sparse. ID-based methods struggle with negative sampling from interaction data alone, lacking semantic information to reflect user preferences, which hampers performance accurately. Following the approach in \cite{yu2022graph}, we categorize the test set into three subsets based on item popularity: Unpopular (the 80\% of items with the fewest clicks), Popular (the top 5\% of items), and Normal (the remaining items). As illustrated in Figure \ref{fig: combined_analysis} (b), HNLMRec significantly outperforms other models in the long-tail subset, primarily due to the enhanced characterization of user preferences through LLM-provided semantic information. In the Popular subset, LightGCN with random negative sampling excels, reflecting a bias toward popular items. Additionally, we further validate the potential of language models in long-tail recommendations from the LLM-driven semantic negative sampling perspective \cite{liu2024large,wu2024coral,liu2024llm}.

\begin{table*}[htbp]
\centering
\caption{
The table below compares the performance of various competing methods and HNLMRec across two datasets. Bold text indicates the highest score, while underlining denotes the best result among the baseline methods. "*" signifies a statistically significant improvement over the best baseline method (i.e., t-test with p < 0.05).}\vspace{-0.3cm}
\setlength\tabcolsep{2pt} 
\label{tab: comparison_main}
\begin{tabular}{cc|cccc|cccc}
\toprule
\hline
\multicolumn{2}{c|}{\multirow{2}{*}{\textbf{Backbone Models}}}& \multicolumn{4}{c|}{\textbf{Toys \& Games}} &\multicolumn{4}{c}{\textbf{CDs \& Vinyl}} \\
\cline{3-10}
 & & \textbf{R@10} & \textbf{N@10} & \textbf{R@20} & \textbf{N@20} & \textbf{R@10} & \textbf{N@10} & \textbf{R@20} & \textbf{N@20}  \\
\hline
\multirow{9}{*}{\textbf{NGCF}} & -RNS & 1.66 & 1.09 & 2.83 & 1.35 & 2.25 & 1.60 & 4.91 & 2.30  \\
 & -DNS & 1.95 & 1.11 & 2.96 & 1.46 & 2.53 & \underline{2.00} & 5.21 & 2.54  \\
 & -MixGCF & 2.14 & \underline{1.39} & 3.12 & 1.61 & 2.57 & 1.91 & \underline{5.37} & 2.56 \\
 & -AHNS & 2.00 & 1.25 & 3.00 & 1.50 & 2.55 & 1.95 & 5.25 & 2.55 \\
 & -KAR & 2.05 & 1.30 & 3.05 & 1.55 & 2.56 & 1.93 & 5.30 & 2.58 \\
 & -LLMRec & 2.08 & 1.20 & 3.15 & 1.60 & 2.60 & 1.92 & 5.20 & 2.57  \\
 & -RLMRec & 1.92 & 1.32 & 3.18 & \underline{1.62} & \underline{2.62} & 1.98 & 5.30 & \underline{2.59}\\
 & -AlphaRec  & \underline{2.19} & 1.38 & \underline{3.19} & 1.61 & \underline{2.62} & 1.96 & 5.33 & 2.57 \\
 \rowcolor{blue!10} & \textbf{OURS} &  \textbf{2.27*} & \textbf{1.46*} & \textbf{3.28*} & \textbf{1.72*} & \textbf{2.73*} & \textbf{2.08*} & \textbf{5.41*} & \textbf{2.65*} \\
\hline
\bottomrule
\vspace{-0.8cm}
\end{tabular}
\end{table*}

\begin{table*}[htbp]
\centering
\caption{
The table below compares the performance of various competing methods and HNLMRec on the Yelp2018 dataset. Bold text indicates the highest score, while underlining denotes the best result among the baseline methods. "*" signifies a statistically significant improvement over the best baseline method (i.e., t-test with p < 0.05).}\vspace{-0.3cm}
\setlength\tabcolsep{2pt} 
\label{tab:yelp_comparison}
\begin{tabular}{cc|cccc|cccc}
\toprule
\hline
\multicolumn{2}{c|}{\multirow{2}{*}{\textbf{Backbone Models}}}& \multicolumn{4}{c|}{\textbf{MF}} &\multicolumn{4}{c}{\textbf{LightGCN}}\\
\cline{3-10}
& & \textbf{R@10} & \textbf{N@10} & \textbf{R@20} & \textbf{N@20} & \textbf{R@10} & \textbf{N@10} & \textbf{R@20} & \textbf{N@20} \\
\hline
\multirow{9}{*}{\textbf{Yelp}} & -RNS & 3.98 & 2.72 & 6.27 & 3.48 & 4.34 & 3.03 & 6.43 & 3.51 \\
 & -DNS & 3.76 & 2.69 & 6.40 & 3.55 & 3.80 & 2.67 & 7.46 & 4.02 \\
 & -MixGCF & 3.61 & 2.58 & 6.94 & 3.63 & 4.40 & 3.06 & 7.68 & \underline{4.11} \\
 & -AHNS & 3.98 & 2.68 & 6.68 & 3.55 & 3.64 & 2.59 & 7.73 & 4.02 \\
 & -KAR & 3.97 & 2.67 & 6.67 & 3.54 & 3.63 & 2.58 & 7.80 & 4.01 \\
 & -LLMRec & 3.62 & 2.59 & 6.68 & 3.55 & 3.62 & 2.57 & 7.69 & 4.00 \\
 & -RLMRec & \underline{4.18} & \underline{2.88} & 7.05 & 3.79 & 4.50 & 3.11 & 7.81 & 4.09 \\
 & -AlphaRec & 4.14 & 2.83 & \underline{7.20} & \underline{3.82} & \underline{4.52} & \underline{3.14} & \underline{7.83} & 4.10 \\
\rowcolor{blue!10} & \textbf{OURS} & \textbf{4.33*} & \textbf{2.96*} & \textbf{7.35*} & \textbf{3.94*} & \textbf{4.66*} & \textbf{3.18*} & \textbf{8.10*} & \textbf{4.25*} \\
\hline
\bottomrule
\end{tabular}
\end{table*}

\begin{algorithm*}[!t]
\caption{Pseudo-code for LLM-Driven Hard Negative Sampling.}
\label{alg: llm-driven-sampling}
\begin{algorithmic}[1]
\Require User-item interaction pair $(u, v)$, item text description $\mathcal{D}_v$, item system prompt $S_v$, user input prompt $\mathcal{D}_u$, user system prompt $S_u$, and hyperparameters $\lambda_1$ and $\lambda_2$.
\Ensure The well-trained model parameters $\Theta$.
\State $\mathcal{P}_v \gets$ Generate the item profile according to Eq.~(\ref{eq: item_profile}).
\State $\mathcal{P}_u \gets$ Generate the user profile according to Eq.~(\ref{eq: user_profile}).
\For{each user-item pair $(u, v)$}
  \State $Neg_i \gets$ Sample semantic hard negatives according to Eq.~(\ref{eq: sematic_sampling}).
\EndFor
\State Transfer contextual information into a fixed-length vector by Eq.~(\ref{eq: trans_to_emb}).
\State Map the vector into latent space by Eq.~(\ref{eq: map}).
\While{not converged}
  \State Compute $\mathcal{L}_{\mathrm{BPR}}$ according to Eq.~(\ref{eq: BPR}).
  \State Compute $\mathcal{L}_{\mathrm{align}}$ for semantic alignment according to Eq.~(\ref{eq: align}).
  \State $\mathcal{L}_{\Theta} \gets \mathcal{L}_{\mathrm{BPR}} + \lambda_1 \cdot \mathcal{L}_{\mathrm{align}} + \lambda_2 \cdot \|\Theta\|_F^2$.
  \State Update parameters $\Theta$ (e.g., by SGD/Adam).
\EndWhile
\State \Return $\Theta$
\end{algorithmic}
\end{algorithm*}

\definecolor{myboxcolor2}{RGB}{191,128,191}
\label{prompt}
\newtcolorbox{mybox3}[1]{
  colback=myboxcolor2!5!white,
  colframe=myboxcolor2!75!black,
  fonttitle=\bfseries,
  title=#1,
  halign title=center, 
  halign=flush center,    
  left=0.5mm,
  right=0.5mm
}
\begin{figure*}[htbp] 
    \centering
\begin{mybox3}{Prompt $\mathcal{S}_v$: Item Profile Generation }
\textbf{System Prompt}:
You will serve as an assistant to help me summarize which types of users would enjoy a specific business.
I will provide you with the basic information (name, city, and category) of that business and also some user feedback for it.
Here are the instructions:\newline
1. The basic information will be described in JSON format, with the following attributes:
\{
    "{\color{blue}name}": "the name of the business",
    "{\color{blue}city}": "city where the company is located", (if there is no city, I will set this value to "None")
    "{\color{blue}categories}": "several tags describing the business" (if there is no categories, I will set this value to "None")
\}\newline
2. Feedback from users will be managed in the following List format:
[
    "{\color{blue}the first feedback}",
    "{\color{blue}the second feedback}",
    "{\color{blue}the third feedback}",
    ....
]

\textbf{Response}: Please provide the answer in JSON format, following this structure:
\{
    "{\color{blue}summarization}": "A summarization of what types of users would enjoy this business" (If unable to summarize, set this value to 'None'.)
    "{\color{blue}reasoning}": "Briefly explain your reasoning for the summarization."
\}
\end{mybox3}
\end{figure*}

\begin{figure*}[htbp] 
    \centering
\begin{mybox3}{Prompt $\mathcal{S}_u$: User Profile Generation }
\textbf{System Prompt}:
You will serve as an assistant to help me determine which types of business a specific user is likely to enjoy.
I will provide you with information about businesses with which the user has interacted and his or her reviews of those businesses.
Here are the instructions:\newline
\textbf{1.} Each interacted business will be described in JSON format, with the following attributes:
\{
    "{\color{blue}title}": "the name of the business", (if there is no business, I will set this value to "None")
    "{\color{blue}description}": "a description of what types of users will like this business",
    "{\color{blue}review}": "the user's review on the business" (if there is no review, I will set this value to "None")
\}
 \newline
\textbf{2.} The information I will give you:
PURCHASED BUSINESSES: a list of JSON strings describing the businesses that the user has interacted.

\textbf{Response}:
Please provide your answer in JSON format, following this structure:
\{
    "{\color{blue}summarization}": "A summarization of what types of business this user is likely to enjoy" (If unable to summarize, set this value to 'None'.)
    "{\color{blue}reasoning}": "Briefly explain your reasoning for the summarization".
\}

\end{mybox3}
\end{figure*}

\begin{figure*}[htbp] 
    \centering
\begin{mybox3}{Prompt $\mathcal{S}_H$: Hard Negative Sample Generation }
\label{prompt: S_H}
textbf{System Prompt}:
You will act as an assistant to help me generate a hard negative sample for a user. 
\textbf{Hard negative samples that are very similar to the user's historical preferences or interaction records but are actually not of interest to the user or do not meet the user's needs}.
Below are the instructions: \newline
\textbf{1.}  User information will be described in JSON format, containing the following attributes:
Each interacted business will be described in JSON format, with the following attributes:
\{
    "{\color{blue}title}": "the name of the business", (if there is no business, I will set this value to "None")
    "{\color{blue}description}": "a description of what types of users will like this business",
    "{\color{blue}review}": "the user's review on the business" (if there is no review, I will set this value to "None")
\}

\textbf{Response}:
Please provide your answer in JSON format, following this structure:
\{
    "hard negative item": "The name of the generated negative sample" 
    (if unable to generate,
     set this value to "None"),
    "{\color{blue}reasoning}": "Briefly explain your reasoning for the summarization".
\}

\end{mybox3}
\end{figure*}

\end{document}